\documentclass[prc,twocolumn,nofootinbib,showpacs,amsmath,superscriptaddress,amssymb,floatfix]{revtex4-1}
\usepackage{dcolumn}
\usepackage{bm}
\usepackage{color}
\usepackage{amssymb}
\usepackage{amsmath}
\usepackage{graphicx}
\usepackage{amsfonts}
\usepackage{slashed}
\usepackage{pstricks}
\usepackage{float}
\usepackage{hyperref}
\usepackage{array}
\allowdisplaybreaks
\usepackage{dcolumn}
\usepackage{epsf}
\usepackage{isotope}
\def\apj{ApJ}

\begin{document}
\title{Impact of three-nucleon forces on gravitational wave emission from neutron stars}
\author{Lucas Tonetto}
\affiliation{Dipartimento di Fisica, ``Sapienza'' University of Rome, Piazzale A. Moro, 5. 00185 Roma, Italy}
\affiliation{INFN, Sezione di Roma, Piazzale A. Moro, 5. 00185 Roma, Italy}
\author{Andrea Sabatucci}
\affiliation{Dipartimento di Fisica, ``Sapienza'' University of Rome, Piazzale A. Moro, 5. 00185 Roma, Italy}
\affiliation{INFN, Sezione di Roma, Piazzale A. Moro, 5. 00185 Roma, Italy}
\author{Omar Benhar}
\affiliation{INFN, Sezione di Roma, Piazzale A. Moro, 5. 00185 Roma, Italy}
\affiliation{Dipartimento di Fisica, ``Sapienza'' University of Rome, Piazzale A. Moro, 5. 00185 Roma, Italy}

\date{\today}

\begin{abstract}
The detection of gravitational radiation, emitted in the aftermath of the excitation of neutron star quasi-normal modes,
has the potential to provide unprecedented access to the properties of matter in the star interior, and shed new light on the
dynamics of nuclear interactions at microscopic level. Of great importance, in this context, will be the sensitivity to the
modelling of three-nucleon interactions, which are known to play a critical role in the high-density regime.
We report the results of a calculation of the frequencies and damping times of the fundamental mode, carried out
using the equation of state of  Akmal, Pandharipande and Ravenhall as a baseline, and varying the strength of the
isoscalar repulsive term the Urbana IX potential within a range consistent with multimessenger astrophysical observations.
The results of our analysis indicate that repulsive three-nucleon interactions strongly affect the stiffness of the
equation of state, which in turn determines the pattern of the gravitational radiation frequencies, largely independent of the mass of the source.
The observational implications are also discussed.

\end{abstract} 


\index{}\maketitle

Ever since their discovery, more than 50 years ago, neutron stars (NSs) have been the subject of extensive theoretical and experimental investigation, aimed at exploiting the potential of these systems
as unparalleled astrophysical laboratories.
The information extracted from the available data, including precise measurements of NS masses~\cite{cromartie2020}
and, more recently, radii~\cite{abbott2017b} and tidal deformabilities~\cite{Riley_2019} have been effectively exploited to test the existing theoretical models.

A primary goal of NS studies is  obtaining information on the Equation of State (EOS) of nuclear matter in the
region of supranuclear densities, in which the microscopic dynamical models are largely unconstrained by the data collected in terrestrial laboratories.
The EOS\textemdash that is, the non trivial relation linking pressure and energy density\textemdash is still largely unknown. The measured properties
of atomic nuclei, supplemented by the available astrophysical observations, do not allow to unambiguously determine either
the nature of the constituents of NS matter, or their interactions over the relevant range of baryon density,
which extends over many orders of magnitude.

In the coming years, new insight will be made possible by the development of multi-messenger astronomy, which has already brought about impressive progress. The combination of  data collected by gravitational-wave (GW) interferometers and
electromagnetic (EM) observatories will provide a powerful tool to advance our understanding of the structure and dynamics of dense nuclear
matter.

The detection of gravitational radiation emitted in the aftermath of the excitation of
Quasi Normal Modes (QNM) may play a most important role, ushering the long anticipated era of neutron star seismology~\cite{AK,astero,leonardo,Anderssonreview}.
The measurement of the frequencies and damping times of NS oscillations, which have been shown to depend significantly on both
the composition and dynamics of matter in the star interior~\cite{astero,BBF}, has the potential to provide unprecedented access not only to the nuclear matter EOS\textemdash whose density dependence is often parametrised in terms of average nuclear properties
such as the compressibility module and the symmetry energy\textemdash  but also to specific  features of the underlying dynamics at microscopic level.

In this article, we report the results of a calculation of the frequencies and damping times of the fundamental mode, routinely referred to as $f$ mode,
carried out using a set of EOS obtained within the framework of non relativistic many-body theory.

The $f$ mode is prominent amongst all possible oscillation modes, because it is a very efficient GW emitter involved in a variety of astrophysical
scenarios, including isolated NS as well as inspiralling binary systems. In isolated NSs, $f$ modes can be excited in core-collapse supernovae leading to the NS formation~\cite{morozowa}, or in a starquake~\cite{KJ:MNRAS}.
In binaries systems, on the other hand, tidal effects include two contributions: a static response\textemdash related to the well known tidal deformability parameter $\Lambda$\textemdash during the adiabatic inspiral, and a resonance between the tidal field and the NS oscillation modes, mostly  the
$f$-mode~\cite{hinderer:nature}.
The emission of gravitational radiation following the excitation of the $f$ mode is also expected to occur in the merger and post-merger phases~\cite{Bauswein,Takami}.
In these regimes, spacetime perturbations must be calculated with respect to a rapidly evolving background, but numerical simulations indicate the
presence of high-frequency components related to the $f$ mode \cite{stergioulas2011,vretinaris2020}. In addition, it has been shown that pre- and
post-merger features can be related~\cite{chakravarti2020, Lioutas2021,bernuzzi2015}. The authors of Ref.~\cite{Lioutas2021} have found an analytic mapping between the $f$-mode frequency of static stars and the dominant GW frequency of merger remnants, allowing to connect the masses in the
two regimes.

In order to investigate the sensitivity to the
description of nuclear dynamics, the Hamiltonian employed to obtain the EOS of Akmal, Pandharipande and Ravenhall (APR)~\cite{APR1998}\textemdash
widely used in neutron star applications\textemdash  has
been modified, changing the value of the parameter $\alpha$, determining the strength of repulsive 
three-nucleon interactions. These interactions are known to be largely isoscalar, and play a dominant role in the high-density region.

The present work can be seen as a complementary follow up to the pioneering study of Ref.~\cite{MSB2021},
whose authors inferred a range of values of $\alpha$ using multi-messenger astrophysical data.
The datasets employed in the analysis of Ref.~\cite{MSB2021} included the GW observation
of the binary NS event GW170817~\cite{abbott2017b}, the
spectroscopic observation of the millisecond pulsars PSR J0030+0451 performed by the NICER
satellite~\cite{Riley_2019}, and the high-precision measurement of the radio pulsars timing of the binary
PSR J0740+6620~\cite{cromartie2020}, providing information on the maximum NS mass that must be supported by the EOS.


This article is organised as follows.
The nuclear Hamiltonian employed in our study, the parametrisation of the three-nucleon potential and the
main features of the corresponding EOSs of dense matter are discussed in Sect.~\ref{sec:3BF}. In Sect.~\ref{subsec:NRO} we outline
the formalism of stellar perturbation theory, employed to obtain the frequencies of QNMs, while the numerical results of our work
are reported in Sect.~\ref{sec:results}. Finally, In Sect.~\ref{sec:conclusion} we summarise our findings and state the conclusions.

\section{Dynamics of neutron star matter} \label{sec:3BF}

Our work is based on the description of nuclear matter as a collection of point like nucleons, whose interactions are described by the
non relativistic Hamiltonian
\begin{equation}
\label{hamiltonian}
H=\sum_{i}\frac{p_i^2}{2m} + \sum_{i<j}v_{ij}+\sum_{i<j<k}V_{ijk} \ ,
\end{equation}
where $m$ and ${\bf p}$ denote the nucleon mass and momentum, respectively.

The nucleon-nucleon (NN) potential $v_{ij}$ is designed to account for the observed properties of the two-nucleon system, in both
bound and scattering states, while inclusion of the three-nucleon (NNN) potential $V_{ijk}$\textemdash required to take into
account processes involving the internal structure of the nucleons~\cite{Friar}\textemdash allows to reproduce
the measured ground-state energy of the NNN bound state and explain saturation of isospin-symmetric
matter (SNM).
The predictive power of the approach based on the Hamiltonian of Eq.~\eqref{hamiltonian} has been firmly established
by the results of calculations carried out using Quantum Monte Carlo techniques,  providing
an accurate description of a variety of properties of nuclei with mass number $A\leq12$~\cite{QMC}.

n this study we have adopted as a baseline the Hamiltonian described in Refs.~\cite{AP1997,APR1998}, comprising the
Argonne $v_{18}$ NN potential (AV18)~\cite{AV18}\textemdash corrected to take into account relativistic boost
corrections, needed to describe NN interactions in the locally inertial frame
associated with a NS\textemdash and the Urbana IX NNN potential  (UIX)~\cite{UIX}.

\subsection{Parametrisation of the NNN potential} \label{NNN:int}

Commonly used phenomenological models of irreducible NNN interactions, such as the UIX potential, are split into two parts
according to
\begin{align}
\label{VNNN}
V_{ijk} = V^{2\pi}_{ijk} +  V^{R}_{ijk} \ .
\end{align}
In the above equation, $V^{2\pi}_{ijk}$ is the attractive Fujita-Miyiazawa potential~\cite{Fujita}, describing
two-pion exchange processes in which a NN interaction leads to the excitation of a $\Delta$ resonance,
while $V^{R}_{ijk}$ is a purely phenomenological repulsive potential.
The Fujita-Miyiazawa mechanism is schematically illustrated in Fig.~\ref{VNNN}.

\begin{figure}[ht]
\includegraphics[scale=0.45]{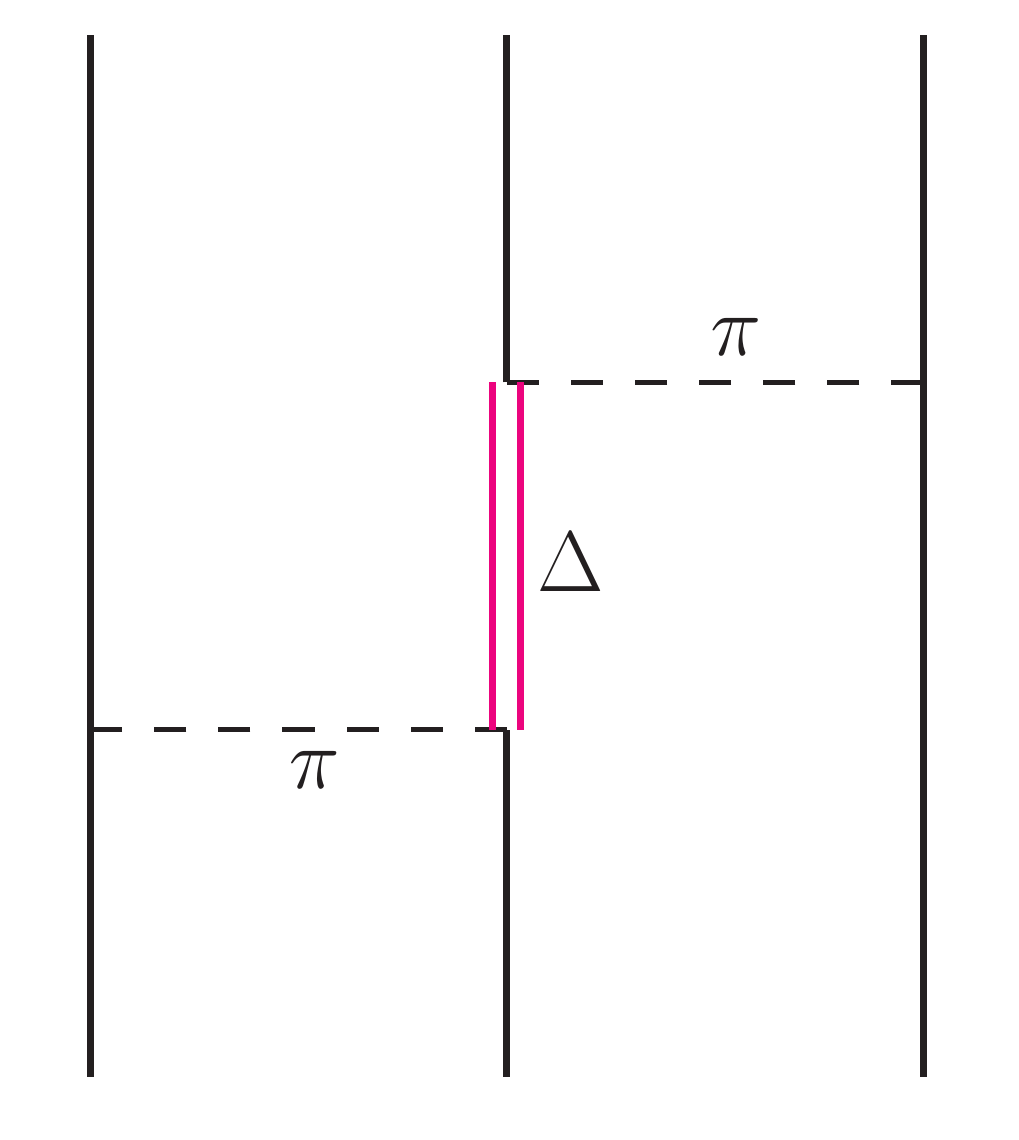}
\vspace*{-.1in}
\caption{Diagrammatic representation of the NNN interaction described by the Fujita-Miyiazawa potential. Dashed lines depict pion-exchange, while
the solid lines and the double line represent nucleons and the $\Delta$ resonance, respectively.}
\label{VNNN}
\end{figure}

The strengths of  $V^{2\pi}_{ijk}$ and $V^{R}_{ijk}$ are adjusted in such a way as to
reproduce the binding energy of \isotope[3][]{He} and  the empirical equilibrium density of  isospin-symmetric matter (SNM),
respectively. On the other hand, the NNN potential is totally unconstrained in the region of
supranuclear densities, relevant to the description of NS properties, in which its contribution becomes important or even dominant.

The contributions of $V^{2\pi}_{ijk}$ and $V^{R}_{ijk}$
to the ground-state energy per nucleon of SNM and pure neutron matter (PNM), obtained from the results reported in
Refs.~\cite{AP1997,APR1998}, are listed in Table~\ref{energies}. It clearly appears that $V^R_{ijk}$ plays the leading role at
$\varrho > \varrho_0$, with $\varrho_0=0.16 \ {\rm fm}^{-3} $ being the equilibrium density of SNM. In addition, owing to its isoscalar nature,
the repulsive potential makes comparable contributions in SNM and PNM.

\begin{center}
\begin{table}[ht]
\begin{tabular}{|c|cc|cc|}
\hline
\hline
\multicolumn{1}{|c}{} &  \multicolumn{2}{c}{$\langle V^R \rangle/A$} & \multicolumn{2}{c|}{$\langle V^{2\pi} \rangle/A$} \\[3pt]
        \hline
        $\varrho $ & \ \ SNM \ \ & \ \ PNM \ \ &  \ \ SNM \ \ & \ \ PNM \ \  \\[1pt]
\multicolumn{1}{|c|}{$[{\rm fm}^{-3}]$}   &  \multicolumn{2}{c|}{$[{\rm MeV]}$} & \multicolumn{2}{c|}{$[{\rm MeV}$]}  \\[3pt]
        \hline 
           &. & & & \\[-8pt]
       0.04 &     0.21 &     0.07 &    -0.36 &     0.08 \\[1pt]
       0.08 &     0.94 &     0.49 &    -0.84 &     0.20 \\[1pt]
       0.12 &     2.15 &     1.38 &    -2.07 &     0.60 \\[1pt]
       0.16 &     4.04 &     2.81 &    -3.64 &     1.23 \\[1pt]
       0.20 &     6.65 &     6.36 &    -5.50 &    -8.67 \\[1pt]
       0.24 &    10.04 &     9.58 &    -8.11 &   -10.07 \\[1pt]
       0.32 &    19.44 &    19.28 &   -13.26 &   -17.39 \\[1pt]
       0.40 &    33.38 &    32.04 &   -38.44 &   -24.22 \\[1pt]
       0.48 &    50.58 &    49.16 &   -50.17 &   -34.09 \\[1pt]
       0.56 &    72.11 &    70.97 &   -65.20 &   -47.19 \\[1pt]
       0.64 &    98.19 &   100.99 &   -81.98 &   -76.88 \\[1pt]
       0.80 &   163.92 &   168.75 &  -117.74 &  -116.32 \\[1pt]
       0.96 &   250.54 &   253.58 &  -169.37 &  -155.01 \\[1pt]
        \hline
        \hline
\end{tabular}
\vspace*{.1in}
\caption{\label{energies}Contributions of $V^{R}_{ijk}$ and $V^{2\pi}_{ijk}$, defined in Eq.~\eqref{VNNN},
to the ground-state energy per nucleon of SNM and PNM at density $\varrho$.
Results adapted from Refs.~\cite{AP1997,APR1998}.}
\end{table}
\end{center}

In the pioneering work of Ref.~\cite{MSB2021}, the authors have modified the strength of the repulsive NNN potential
through the replacement
\begin{align}
\label{modification}
V^{R}  \to \alpha \  V^{R}  \  ,
\end{align}
leading to  a sizeable modification of the nuclear matter EOS at $\varrho > \varrho_0$, and studied  the $\alpha$-dependence of
NS properties such as the mass-radius relation and the tidal deformation. The ultimate goal of this analysis was exploring
the potential to infer a constraint on the value of $\alpha$ from available and upcoming astrophysical data.

In this article, we report the results of a similar and complementary study, aimed at pinning down the footprint of repulsive NNN interactions on the gravitational radiation associated with excitation of the fundamental mode of NSs.

We employ the parametrisation of the energy density of nuclear matter at baryon density
$\varrho$ and proton fraction $x$ discussed in Ref.~\cite{APR1998}, yielding
an accurate description of the PNM and SNM energies obtained by Akmal, Pandharipande and Ravenhall using a variational approach and the AV18+UIX Hamiltonian~\cite{AP1997}.
The explicit expression
\begin{align}
\label{energy_fit1}
\epsilon(\varrho,x) & = \left[ \frac{1}{2m}+f(\varrho,x) \right]\tau_p \\
\nonumber  & + \left[\frac{1}{2m}+f(\varrho,1-x)\right]\tau_n+g(\varrho,x),
\end{align}
with
\begin{align}
\tau_p = \frac{1}{5 \pi^2} \left( 3 \pi^2 x \varrho \right)^{5/3} \  ,
\end{align}
\begin{align}
\tau_n = \frac{1}{5 \pi^2} \left[ 3 \pi^2 (1-x)\varrho \right]^{5/3}  ,
\end{align}
and
\begin{align}
g(\varrho,x) & = g(\varrho,1/2)\left[1-(1-2x)^2\right] \\
\nonumber
& +  g(\varrho,0)(1-2x)^2 ,
\end{align}
involves 21 parameters.
The first two terms in the right hand side of Eq.(\ref{energy_fit1}) account for the kinetic contribution to the energy density, while the function
$g(\varrho,x)$ describes the interaction energy. The explicit form of the functions $f(\varrho,x)$ and $g(\varrho,x)$, as well as the numerical
values of the parameters for both PNM and SNM,  can be found in Ref.~\cite{APR1998}.

For any densities and proton fractions, changing the strength of  $V^R$ according to  Eq.~\eqref{modification} leads to a change of the interaction contribution to the
 energy density
\begin{align}
\delta g(\varrho,x,\alpha) & = \delta g(\varrho,1/2,\alpha)\left[1-(1-2x)^2\right] \\
\nonumber
& +  \delta g(\varrho,0,\alpha)(1-2x)^2 \ ,
\end{align}
which can be computed at first order of perturbation theory using
\begin{align}
 \delta g(\varrho,1/2,\alpha) = \frac{\varrho}{A}  (\alpha - 1) \langle V^R \rangle_{SNM} \ , \\
 \delta g(\varrho,0,\alpha) = \frac{\varrho}{A}  (\alpha - 1) \langle V^R \rangle_{PNM} \ .
 \end{align}
The above equations show that $\delta g(\varrho,x)$ can be readily obtained from the values of  $\langle V^R \rangle$ of Table~\ref{energies},
the density dependence of which is accurately approximated by a polynomial including powers up to $\varrho^3$.

\begin{figure}[ht]
\includegraphics[scale=0.7]{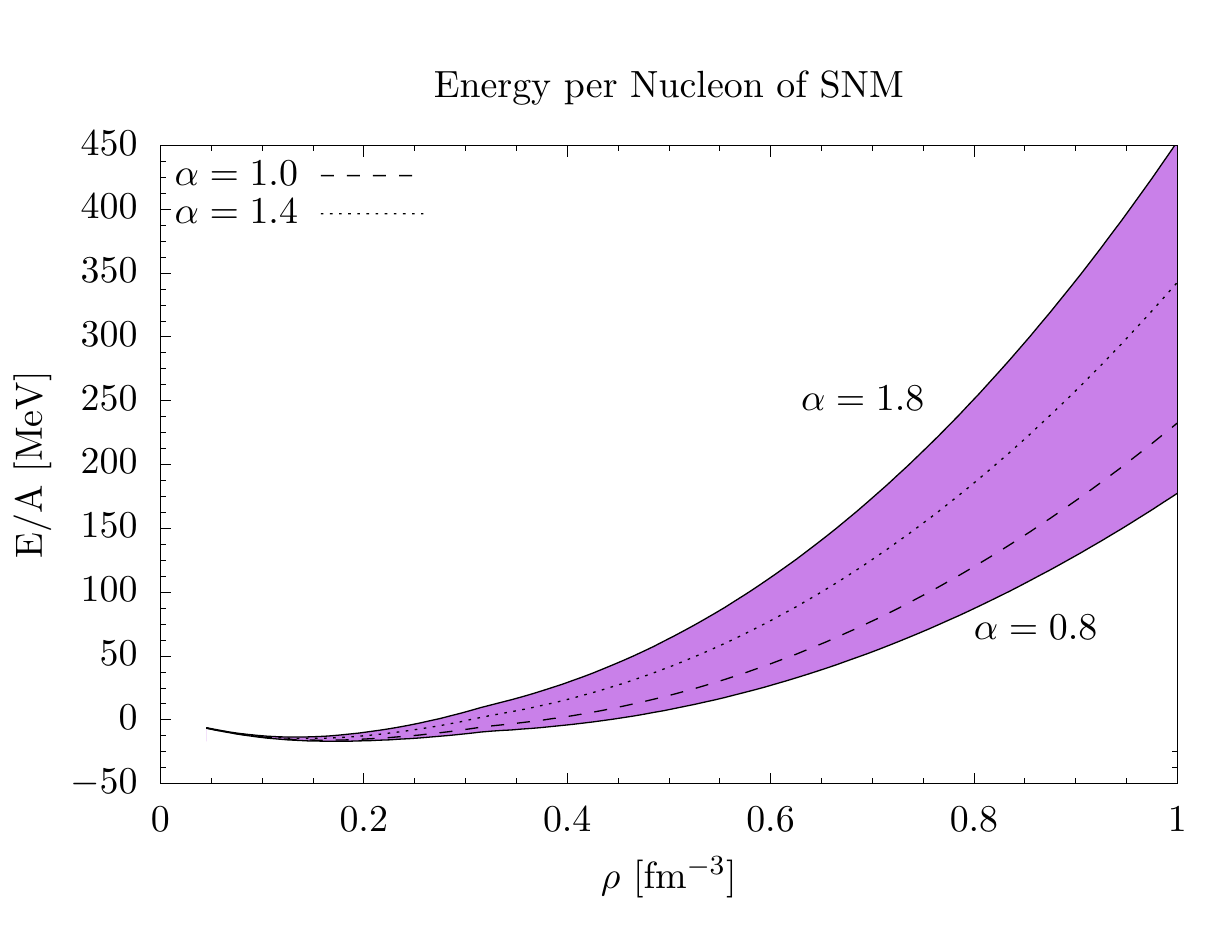}
\vspace*{-.2in}
\caption{Density dependence of the energy per nucleon of SNM computed varying the strength of the repulsive NNN potential $\alpha_R$,
defined by Eq.~\eqref{VNNN}, in the range $0.8 \leq \alpha \leq 1.8$.}
\label{ESNM}
\end{figure}

Figure~\ref{ESNM} shows the energy per nucleon of SNM, displayed as a function of density, for  $\alpha =$ 0.8, 1.0, 1,4, and 1.8.
The range of $\alpha$ has been chosen in such a way as to preserve the equilibrium density of SNM reported in Ref.~\cite{APR1998}, corresponding to $\alpha=1$.

\subsection{Speed of sound} \label{subsec:nucqua}

\begin{figure}[ht]
\includegraphics[scale=0.65]{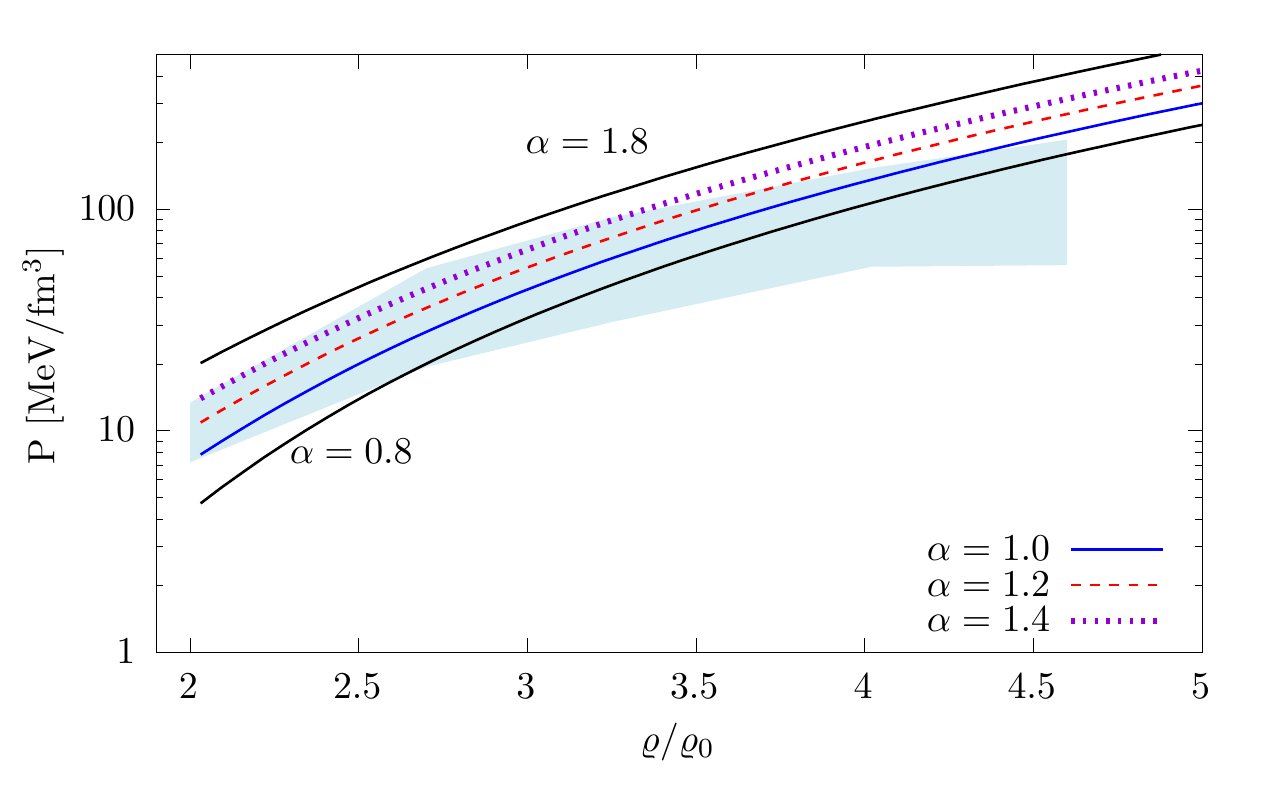}
\vspace*{-.1in}
\caption{Density dependence of the pressure of SNM, obtained using the parametrisation of
$V^R_{ijk}$ of Eq.~\eqref{modification} with $0.8 \leq \alpha \leq 1.8$. The shaded region
represents the results of the analysis of  Ref.~~\cite{danielewicz}.}
\label{pressure}
\end{figure}

The results  of Fig.~\ref{ESNM} clearly show that the strength of the NNN repulsive potential strongly affects the matter pressure
\begin{align}
P = \varrho^2 \frac{\partial (E/A)}{\partial \varrho} \ ,
\end{align}
which in turn determines the speed of sound, $v_s$, defined through\footnote{Note that in this article we adopt the 
system of natural units, in which $\hbar = c = 1$.}
\begin{align}
v^2_s =   \frac{1}{\varrho} \frac{\partial P}{\partial (E/A)}  \ .
\end{align}
In Fig.~\ref{pressure}, the pressure of SNM corresponding to different values of $\alpha$ is displayed as a function of density.
For comparison, the shaded area shows the region consistent with the experimental data discussed in Ref.~\cite{danielewicz},
providing a constraint on $P(\varrho)$ at  $\varrho \geq 2 \varrho_0$.

\begin{figure}[ht]
\includegraphics[scale=0.725]{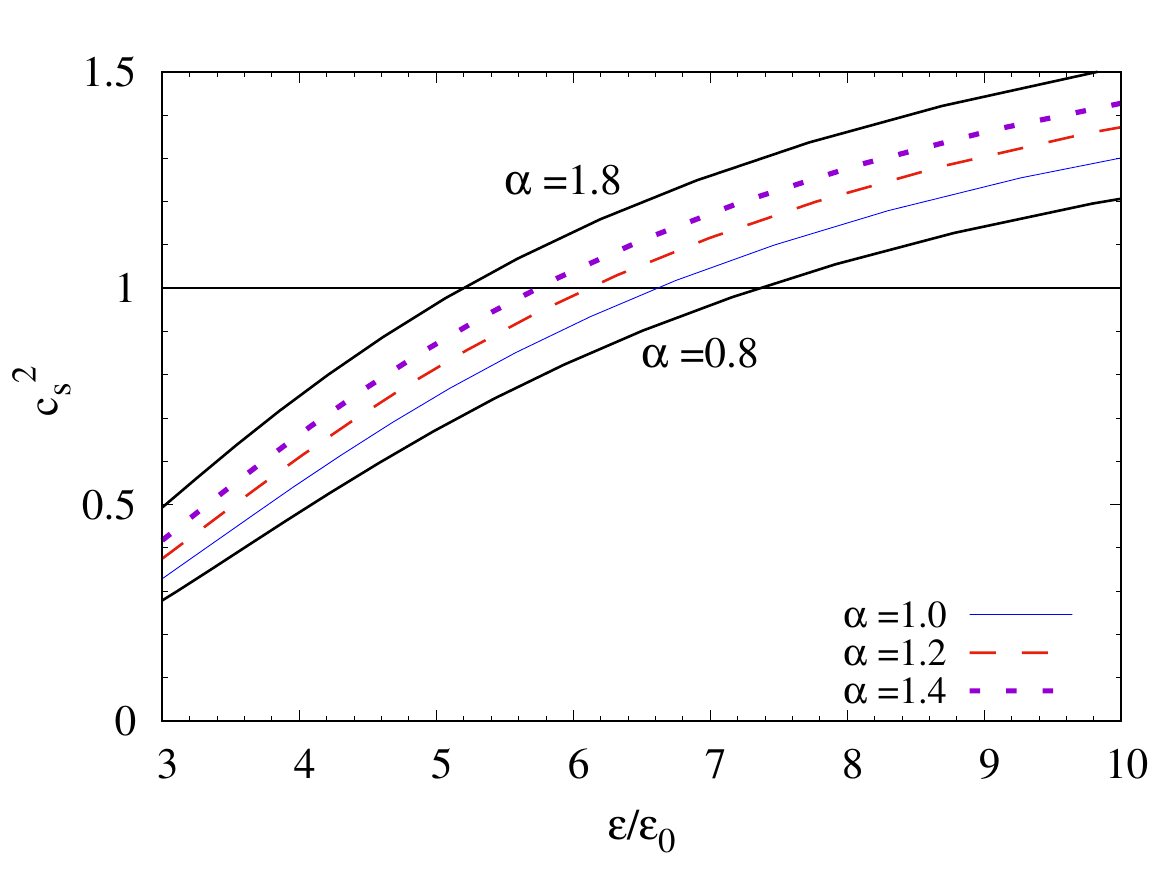}
\vspace*{-.25in}
\caption{Energy density dependence of the squared speed of sound in $\beta$-stable matter, obtained using the parametrisation of
$V^R_{ijk}$ of Eq.~\eqref{modification} with $0.8 \leq \alpha \leq 1.8$.}
\label{speedofsound}
\end{figure}

Being non relativistic in nature, the approach based on the Hamiltonian of Eq.\eqref{hamiltonian} unavoidably leads to a violation of
causality\textemdash signalled by a speed of sound exceeding unity\textemdash in the limit of large density. The results of Fig.~\ref{speedofsound},
obtained using the parametrisation of
$V^R_{ijk}$ of Eq.~\eqref{modification} with $0.8 \leq \alpha \leq 1.8$,
show that in $\beta$-stable matter with $\alpha =1$ the region of $c^2_s > 1$ corresponds  to energy density $\epsilon \gtrsim 6.5 \epsilon_0$, while for larger values of  $\alpha$ the violation of causality is pushed to $\epsilon$ as low as $\sim 4 \epsilon_0$.
Here, $\epsilon_0 = 2.67\times 10^{14} \ {\rm g \ cm}^{-3}$ is the mass density of a system of nucleons at number density $\varrho_0$.

\begin{figure}[ht]
\includegraphics[scale=0.65]{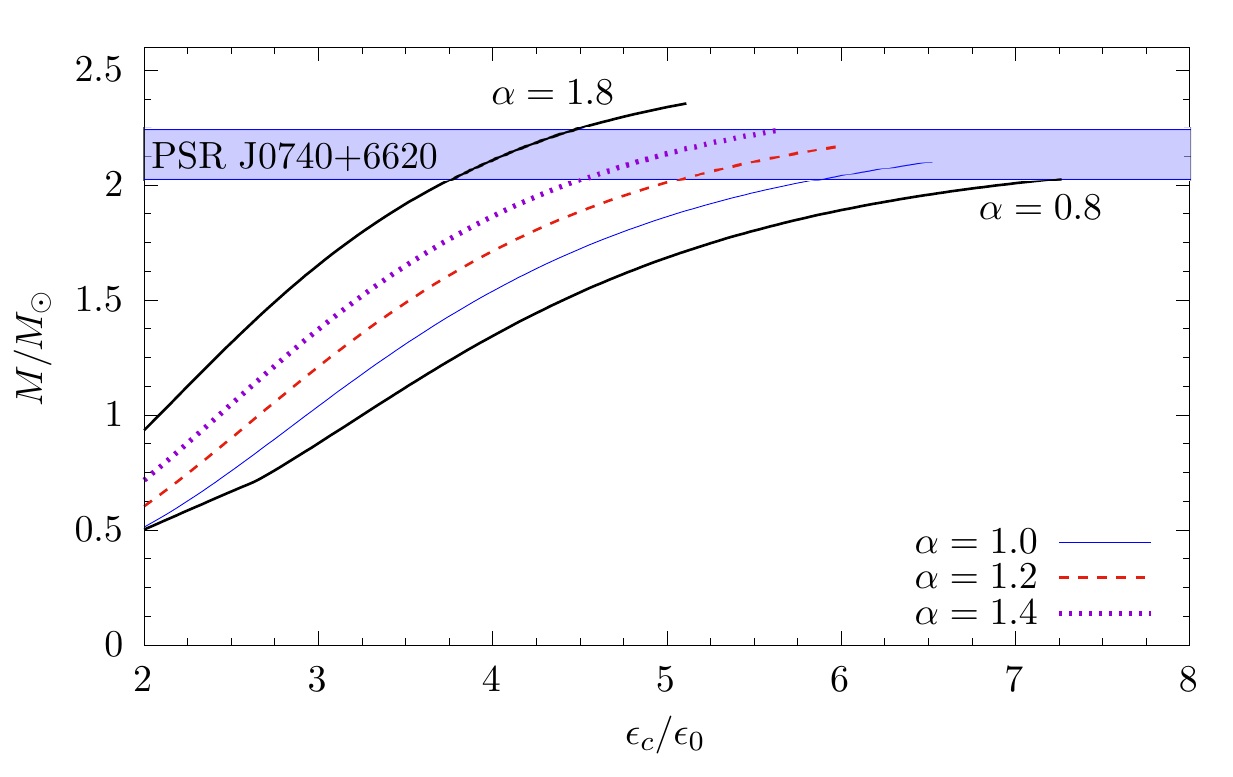}
\vspace*{-.1in}
\caption{Energy density dependence of the NS mass, obtained from the solution of the TOV equations using the parametrisation of
$V^R_{ijk}$ of Eq.~\eqref{modification} with $0.8 \leq \alpha \leq 1.8$. All curves extend to the maximum value of density
for which the speed of sound in matter satisfies the causality constraint $v_s<1$.}
\label{masseps}
\end{figure}

The capability of the dynamical model employed in this work to support a NS of mass
compatible with the observational constraints can be gauged comparing the results of Figs.~\ref{speedofsound} and~\ref{masseps}.
The dependence of the NS mass of\textemdash obtained from the solution of the Tolman-Oppenheimer-Volkoff (TOV)
equations\textemdash on the central energy density, $\epsilon_c$, clearly shows that all EOSs corrresponding to $0.8 \leq \alpha \leq 1.8$ predict
a stable NS with mass exceeding two solar masses, and central densities well below the region corresponding to $c_s^2 >1$.

\section{Frequencies of non-Radial Oscillation Modes}
\label{subsec:NRO}

In this work, we analyse the influence of the strength of the repulsive NNN forces on the gravitational radiation  emitted
by a NS following the excitation of the non-radial $f$ mode~\cite{thornecampolattaro1967,thornecampolattaro1967erratum}.
For any assigned EOS of  NS matter, the frequencies of QNMs can be obtained by studying the
source-free, adiabatic perturbations of an equilibrium configuration. This involves solving the linearised Einstein
equations, coupled with the equations of hydrodynamics,
with suitably posed boundary conditions.

The unperturbed configuration is assumed to be a static spherically symmetric star, whose geometry is described by the
line element
\begin{equation}
(ds^{2})^{(B)}=-e^{2\nu} dt^{2} + e^{2\lambda} dr^{2} + r^{2}(d\theta^{2}+\sin^{2}{\theta}d\phi^{2}),
\label{eq:metric_spherical}
\end{equation}
where $\nu$ and $\lambda$ are functions of $r$, and $\lambda$ is related to the gravitational mass comprised within a sphere of radius
$r$, $M(r)$, through
\begin{equation}
e^{-\lambda}=\left[ 1-2\frac{ M(r)}{ r }\right]^{1/2} \ .
\label{eq:lambda_static}
\end{equation}
Denoting by $\epsilon(r)$ and $p(r)$ the total energy density and pressure measured in the proper frame of the fluid, we can write the Tolman-Oppenheimer-Volkoff (TOV) equations in the form
\begin{align}
\label{eq:TOV1}
   \frac{dM}{dr} & =4 \pi r^2 \epsilon \ , \\
\label{eq:TOV2}
    \frac{dp}{dr} & = -\frac{(\epsilon+p)(M+4 \pi r^3 p)}{r(r-2M)} \ , \\
\label{eq:TOV3}
    \frac{d \nu}{dr} & = - \frac{1}{(\epsilon+p)} \frac{dp}{dr} \ ,
\end{align}
with the boundary conditions
\begin{align}
\label{eq:TOVB1}
  M(0) & = 0 \  \ , \ \  p(R) = 0 \ , \\
\label{eq:TOVB2}
   \nu(R) & = \frac{1}{2} \ln \bigg( 1-\frac{2M(R)}{R} \bigg) \ .
\end{align}

Here, we write the perturbed line element in the form \cite{lindblomdetweiler1985}
\begin{equation}
\begin{aligned}
 ds^{2}= &-e^{2\nu} (1+r^l H_0 Y^l_m e^{i\omega t}) dt^{2}  \\
         & - 2i\omega r^{l+1} H_1 Y^l_m e^{i\omega t} dtdr \\
         & + e^{2\lambda}(1-r^l H_0 Y^l_m e^{i\omega t}) dr^2  \\
         & + r^{2}(1-r^l K Y^l_m e^{i\omega t})(d\theta^{2}+\sin^{2}{\theta}d\phi^{2}),
\label{eq:metric_DL}
\end{aligned}
\end{equation}
where the functions $Y_{lm}(\theta, \phi)$ are spherical harmonics. The small amplitude motion of the perturbed configuration is described by the Lagrangian 3-vector fluid displacement $\xi^j$, which can be represented in terms of perturbation functions $W(r)$ and $V(r)$ as
\begin{eqnarray}
 \xi^r&=&r^{l-1} e^{-\lambda} W(r) Y^l_m e^{i\omega t}, \label{eq:Xi} \\
 \xi^\theta&=&-r^{l-2} V(r) \partial_\theta Y^l_m e^{i\omega t}, \\
 \xi^\phi&=&-\frac{r}{(r \ \sin\theta)^{2} }V(r) \partial_\phi Y^l_m e^{i\omega t}.
\label{eq:lagrangian_3vector}
\end{eqnarray}
Our analysis will be restricted to the $l = 2$ component. Introducing the variable $X$, defined as
\begin{align}
    X & = \omega^2 (\epsilon+p) e^{-\nu} V - r^{-1} p,_r e^{(\nu-\lambda)} W \\
    \nonumber & + \frac{1}{2} (\epsilon+p) e^{\nu} H_0 \ ,
    \label{eq:definitionX}
\end{align}
where the notation $,_r$ indicates differentiation with respect to $r$, it is possible to write a
fourth-order system of linear differential equations for the functions $H_{1}$, $Z$, $W$, and $X$~\cite{lindblomdetweiler1983,lindblomdetweiler1985}

\begin{align}
\nonumber
K,_r & = \frac{H_0}{r} + \frac{l(l+1)}{2r} H_1 - \left[ \frac{ (l+1)}{r}-\nu,_r \right]K \\
& - 8 \pi (\epsilon+p) \frac{e^{\lambda}}{r} W \ ,
\end{align}
\begin{align}
\nonumber
H_1,_r  & = -\frac{1}{r}  \left[ l+1+2M \frac{e^{2\lambda}}{r} + 4 \pi r^2 e^{2\lambda} (p-\epsilon) \right] H_1 \\
            & + \frac{ e^{2 \lambda} }{r} \left[ H_0 + K - 16 \pi (\epsilon+p) V \right] \ ,
\end{align}
\begin{align}
\nonumber
W,_r &= -\frac{ (l+1) }{r}W + r e^{\lambda} \left[ \frac{1}{\gamma p} e^{-\nu} X  \right. \\
& - \left. \frac{l(l+1)}{r^2} V + \frac{1}{2} H_0 + K\right] \ ,
\end{align}
\begin{align}
 \nonumber
X,_r &=  -\frac{l}{ r} X + (\epsilon+p)e^{\nu} \left\{  \frac{1}{2} \left( \frac{1}{r}-\nu,_r \right) H_0  \right.\\
 \nonumber
& + \frac{1}{2} \left[ r \omega^2 e^{-2\nu}  + \frac{1}{2} \frac{ l(l+1)}{r}\right] H_1 + \frac{1}{2} \left( 3 \nu,_r - \frac{1}{r} \right)K  \nonumber \\
 \nonumber
&- \frac{ l(l+1) \nu,_r}{r^{2}} V - \frac{1}{r}  \bigg[ 4 \pi (\epsilon+p) e^{\lambda} \\
& + \left.  \omega^2 e^{\lambda-2\nu}- r^2 \left( \frac{ e^{-\lambda} \nu,_r  } {r^2} \right),_r \bigg] W  \right\} \ .
\end{align}
In the above equations, $\gamma$ is the adiabatic index defined by
\begin{equation}
    \gamma = \frac{(\epsilon+p)}{p}\frac{\Delta p}{\Delta \epsilon},
    \label{eq:adiabaticindex}
\end{equation}
and  $H_0$ can be taken out using
\begin{align}
\nonumber
& \left[ 3M  +\frac{1}{2}(l+2)(l-1)r+4 \pi r^3 p \right] H_0  = 8 \pi r^3 e^{-\nu} X \\
 \nonumber
 & \ \ \ \ \ \ \  - \left[ \frac{1}{2}l(l+1)(M+4 \pi r^3 p) - \omega^2 r^3 e^{-2(\lambda + \nu)} \right] H_1  \\
 & \ \ \ \ \ \ \ +  \bigg[ \frac{1}{2}(l+2)(l-1)r-\omega^2 r^3 e^{-2\nu}   \\
 \nonumber
 & \ \ \ \ \ \ \   - \frac{1}{r} e^{2\lambda} (M+4 \pi r^3 p)(3M - r + 4 \pi r^3 p) \bigg] K  .
\label{eq:relation_variables}
\end{align}

The following boundary conditions need to be satisfied: (i) the perturbation functions must be finite everywhere, particularly
at $r=$0 where the system becomes singular, and (ii) the perturbed pressure must vanish at the surface of the star, corresponding to $r=R$, at any time, implying $\Delta p (R)$=0. From the relation~\cite{sotani2011}
\begin{equation}
    \Delta p = -r^l e^{-\nu} X \ ,
    \label{eq:deltap_X}
\end{equation}
it follows immediately that $\Delta p (R)$=0 also implies $X(R)=$0. For a given set of $l$ and $\omega$, there is a unique solution which satisfies all of the boundary conditions.

To solve the equations numerically we expand the solutions at $r=$0 and $r=R$, according to a procedure suggested by Lindblom and Detweiler \cite{lindblomdetweiler1983,lindblomdetweiler1985} and refined by the authors of Ref.~\cite{lusuen2011}.

Outside the star all quantities associated with the fluid vanish, and the perturbation equations reduce to the Zerilli equation \cite{zerilli1970,fackerell1971,chandrasekhardetweiler1975}
\begin{equation}
    \frac{d^2 Z}{dr^{*2}}+[\omega^2-V(r^*)]Z=0 \ ,
    \label{eq:Zerillieq}
\end{equation}
where the effective potential $V(r^*)$ is written as
\begin{eqnarray}
    V(r^*) &=& \frac{2(1-2M/r)}{r^3 (nr+3M)^2} \Big[ n^2 (n+1)r^3+3n^2 M r^2  \nonumber \\
    && + 9nM^2 r + 9M^3 \Big] \ ,
\end{eqnarray}
with $n=(l-1)(l+2)/2$, and $r^*$ is the ``tortoise'' coordinate, which can be written in terms of $r$ as
\begin{equation}
    r^* = r + 2M \log \left(\frac{r}{2M}-1 \right) \ .
\end{equation}

In terms of $H_0(r)$ and $K(r)$, the Zerilli function $Z(r^*)$ and its derivative are
\begin{align}
    Z(r^*) &= \frac{k(r)K(r)-a(r)H_0(r)-b(r)K(r)}{k(r)g(r)-h(r)}, \label{eq:Zerilli_function} \\
    \frac{dZ(r^*)}{dr^*} &= \frac{h(r)K(r)-a(r)g(r)H_0(r)-b(r)g(r)K(r)}{h(r)-k(r)g(r)}, \label{eq:Zerilli_derivative}
\end{align}
with \cite{lusuen2011}
\begin{align}
    a(r) &= -\frac{ (nr+3M)}{[\omega^2 r^2 - (n+1)M/r]} \ , \\
    b(r) &= \frac{[nr(r-2M)-\omega^2 r^4 + M(r-3M)]}{(r-2M)[\omega^2 r^2 - (n+1)M/r]}, \\
    g(r) &= \frac{[n(n+1)r^2+3nMr+6M^2]}{r^2 (nr+3M)}, \\
    h(r) &= \frac{[-nr^2+3nMr+3M^2]}{(r-2M)(nr+3M)}, \\
    k(r) &= -\frac{r^2}{(r-2M)} \ .
\end{align}
The Zerilli equation has two linearly independent solutions $Z_+ (r^*)$ and $Z_- (r^*)$. They correspond to incoming and outgoing GWs, respectively. The general solution for  $Z(r^*)$ is given by the linear combination
\begin{equation}
    Z(r^*) = A(\omega)Z_- (r^*) + B(\omega)Z_+ (r^*). \label{eq:Zer_expansion}
\end{equation}
At large $r$, $Z_+$ and $Z_-$ can be conveniently expanded according to
\begin{align}
    Z_-(r^*) &= e^{-i \omega r^*} \sum_{j=0}^\infty \beta_j r^{-j}, \label{eq:Zer_minus_infty}\\
    Z_+(r^*) &= e^{i \omega r^*} \sum_{j=0}^\infty \overline{\beta}_j r^{-j}, \label{eq:Zer_plus_infty}
\end{align}
where ${\overline{\beta}}_j$ denotes  the complex conjugate of $\beta_j$. Keeping only terms up to  $j=\mathrm{2}$ and
substituting  Eq.~\eqref{eq:Zer_minus_infty}  into Eq.~\eqref{eq:Zerillieq}, we obtain \cite{lusuen2011}
\begin{align}
    \beta_1 &= -i (n+1) \omega^{-1} \beta_0, \\
    \beta_2 = -\omega^2 & \left[ \frac{1}{2}n(n+1)-\frac{3}{2}iM \omega \left( 1+\frac{2}{n} \right) \right] \beta_0.
\end{align}

As we are interested in purely outgoing radiation, we must find the value of $\omega$ satisfying $B(\omega)=0$,
 which is the frequency of the desired QNM.

\section{Numerical Results} \label{sec:results}

\begin{figure}[ht]
\includegraphics[scale=0.275]{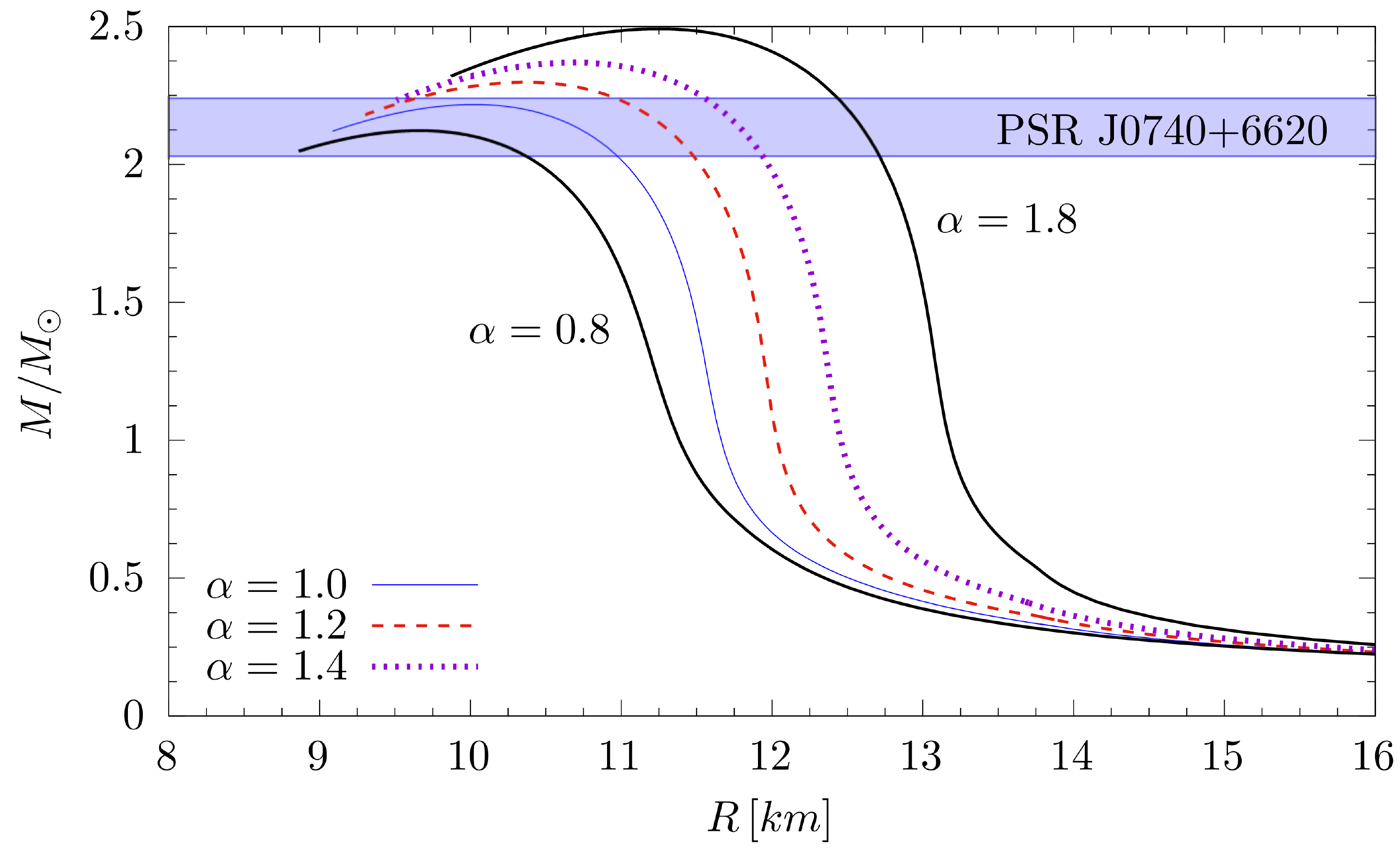}
\caption{Mass radius relations of the stable NS configurations obtained using the EOSs corresponding to the nuclear 
Hamiltonian discussed in Sect.~\ref{sec:3BF}. The curves are labelled according to the value of the parameter $\alpha$,
defined in Sect.~\ref{NNN:int} .}
\label{fig:MR}
\end{figure}

Changing the strength of the repulsive contribution to the NNN potential, $V^R_{ijk}$, obviously affects the stiffness of the EOS of $\beta$-stable matter, see Sect.~\ref{subsec:nucqua}. This effect is clearly illustrated in Fig.~\ref{fig:MR}, showing that higher values of $\alpha$ increase both the stellar maximum mass and the radius. At both $M =  \mathrm{1.4} \ M_\odot$ and $M = \mathrm{2} \ M_\odot$, increasing the value of $\alpha$ from 1 to 1.6 leads to $\sim$ 10 \% increase of
the star radius.

\begin{figure}[htb]
    \centering
     \includegraphics[width=\columnwidth]{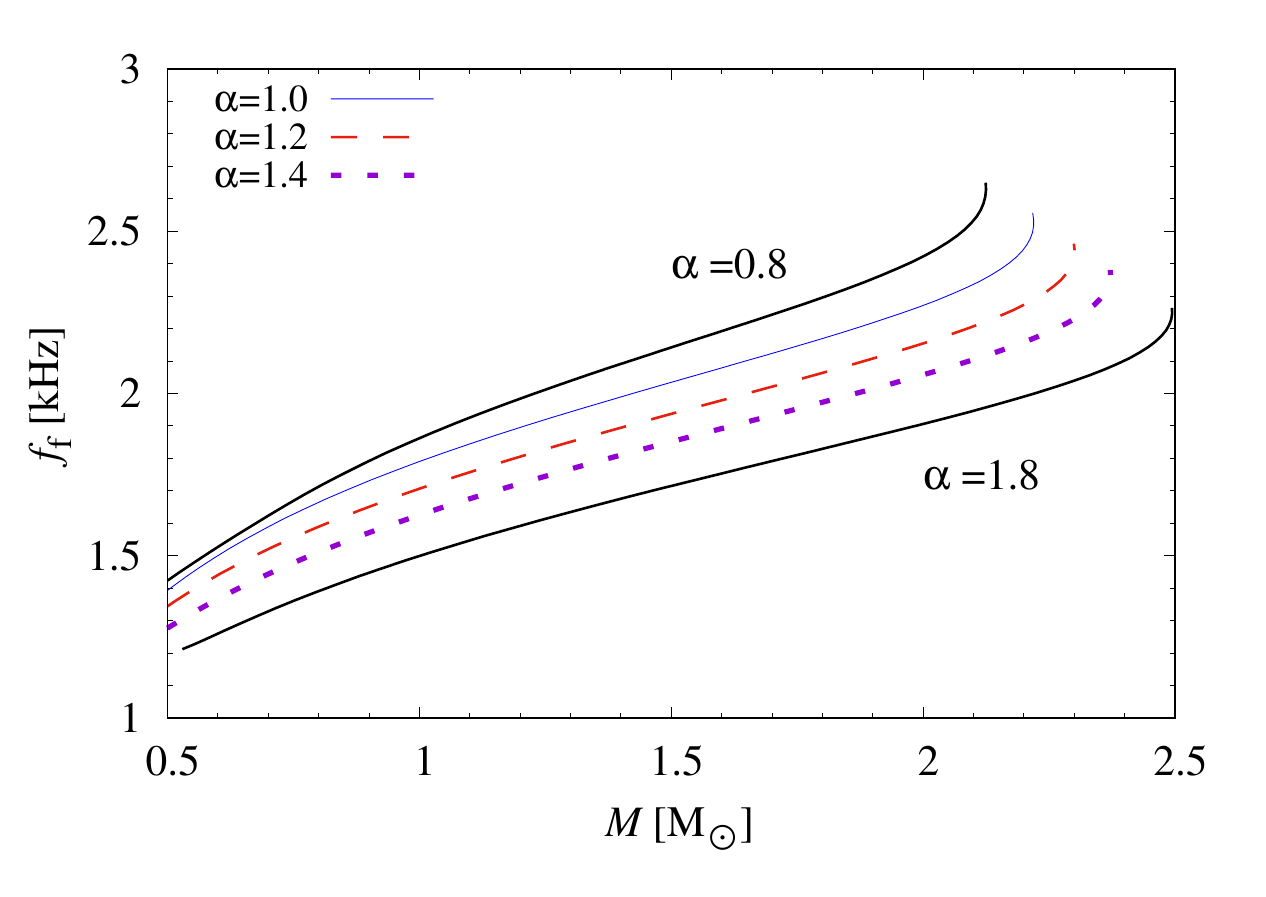}
      \includegraphics[width=\columnwidth]{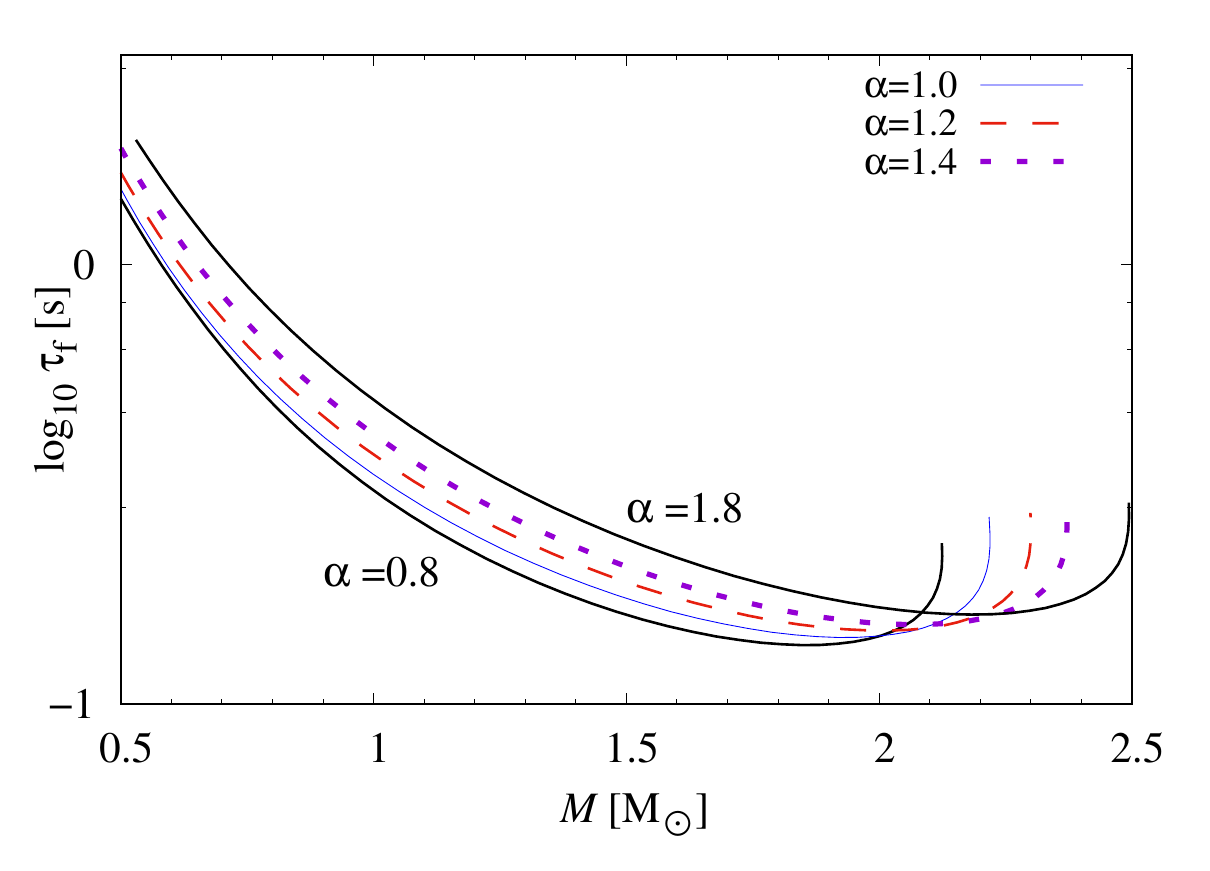} 
      \vspace*{-.25in}
   \caption{Top panel: frequency of the $f$ mode of stable NSs configurations obtained using the nuclear Hamiltonian discussed in Sect.~\ref{NNN:int}.
   The curves are labelled by the value of the parameter $\alpha$, determining the strength of repulsive NNN interactions. Bottom panel: same as the top panel, but for the damping time $\tau=\mathrm{1/Im}\{\omega\}$}
    \label{fig:modes}
\end{figure}

The pattern emerging from Fig.~\ref{fig:MR}, indicating that the mass-radius relations are ordered according the value of $\alpha$\textemdash which in turn determines the stiffness\textemdash is largely reflected by the results displayed in the top panel of Fig.~\ref{fig:modes}, showing the pulsation frequencies of the $f$ mode, $f=\mathrm{Re}\{\omega\}/2\pi$, as a function of the star mass. The different curves correspond to different
values of $\alpha$ in the range $0.8 \leq \alpha \leq 1.8$. Note that the multimessenger analysis of Ref.~\cite{MSB2021}\textemdash in which
the stronger constraint turns out to be  the bound on the maximum mass provided by PSR J0740+6620\textemdash yields a
probability distribution for $\alpha$ that, while being compatible with the value $\alpha=1$, corresponding to the APR EOS, shows large support
for $\alpha >1$, corresponding to stronger NNN repulsion and stiffer EOSs. The role of stiffness in the determination of the tidal polarizability 
$\Lambda$ has been also discussed in Ref.~\cite{Sabatucci2020}, whose authors compared the results of calculations performed using EOSs obtained from different models of nuclear dynamics.


It is apparent that increasing $\alpha$ results in sizeably lower values of the frequency. A comparison between the cases $\alpha=0.8$ and
$\alpha=1.8$ at $M = \mathrm{1.4  \ M_\odot}$  shows a decrease of $\sim 20$\%, from 2.1 kHz to 1.68 kHz. The same percentage decrease, from 2.42 kHz to 1.92 kHz is found at  $M = \mathrm{2  \ M_\odot}$. The mass-independence of the effect of using EOSs featuring different stiffness is a remarkable property, that was already pointed out,  in a somewhat different context,  by the authors of Ref.~\cite{BBF}.

On the other hand, the bottom panel of Fig.~\ref{fig:modes} shows that the damping times $\tau=\mathrm{1/Im}\{\omega\}$ corresponding
to different values of $\alpha$ are much closer to one another. Moreover, they are not ordered according to the stiffness of the underlying EOSs over the whole range of mass.

\section{Summary and Conclusions} \label{sec:conclusion}

We have studied the dependence of the frequencies and damping time of the NSs' $f$ mode on the
strength of repulsive NNN forces. Improving the available NNN potential models is of outmost
importance for astrophysical applications, because NNN interactions, while playing a critical role in the
determination of the nuclear matter EOS at high densities, is not strongly constrained by nuclear data

In our dynamical model, the repulsive contribution to the UIX NNN potential of the APR Hamiltonian
is modified through the action of the parameter $\alpha$. The APR model, providing the baseline
of the analysis, corresponds to $\alpha =1$, and larger values of $\alpha$ lead to
stiffer EOSs. Based on the results of Ref.~\cite{MSB2021}, we have considered the range
$0.8 \leq \alpha \leq 1.8$.

We find that the frequency of the $f$ mode is significantly affected by the strength of the NNN potential over the
whole range of stellar masses, while the damping time turns out to be largely independent of $\alpha$.
These results suggest that, with a galactic supernova event, we may be able to exploit the detection
of gravitational radiation associated with the excitation of the QNM to constrain the value $\alpha$.
However, GW observations coming from mergers may offer a more promising scenario.

The most important way of analysing the influence of QNMs in a merger waveform would be taking into account the effects of dynamical tides \cite{Anderssonreview,hinderer:nature,Steinhoff:2021dsn,schmidthinderer2019,Andersson:2019dwg}. As the binary evolves towards merger, variations in the tidal fields reach a resonance with the star's internal oscillation modes, creating new particular features in the GW waveform that can be detected and provide information on the QNMs; amongst these modes, the $f$ mode is the most important one \cite{hinderer:nature}. Most notably, Ref. \cite{hinderer:nature} has shown promising results that we will be able to measure the $f$-frequency from GW inspirals to within of Hz in the future GW detector networks. At this stage, we would be able to directly infer how strong $V_{ijk}^R$ is, as the differences we get are of hundreds of hertz. As also noted in their work, GW170817 rules out EOSs that produce very low $f$ frequencies, corresponding to very stiff EOSs and higher values of $\alpha$.

Exploring the post-merger signal is another way of gathering information on the $f$ mode of static stars. This scenario consists of a rapidly evolving background spacetime, but even so the peak frequency of GW emission is related to the $f$ mode \cite{stergioulas2011,vretinaris2020} and moreover, we can establish relations between this frequency and the one of static stars \cite{chakravarti2020, Lioutas2021,bernuzzi2015}. In the next generation of GW detectors we would be able to detect the post-merger signal, which in turn would give us the possibility of constraining the $f$ frequency of the inspiral phase, allowing to have another valuable information about the internal composition of NSs.


\acknowledgments
This work has been supported by the Italian National Institute for Nuclear
Research (INFN) under grant TEONGRAV.
%
%

\end{document}